\documentclass[aps,prl,showpacs,reprint]{revtex4-1}

\usepackage{bm}
\usepackage{amsmath}
\usepackage{graphicx}
\usepackage{subfigure}
\usepackage{color}
\usepackage{amsmath}    
\usepackage{color}      
\usepackage{amsfonts}
\usepackage{amssymb}
\usepackage{epsfig}
\usepackage{ifthen}
\usepackage{multirow}
\usepackage{bigstrut}
\usepackage{rotating}
\usepackage{longtable}
\usepackage{color,calc}

\renewcommand{\vec}{\bm}

\def\i{\text{i}}
\def\splitting{\varepsilon}
\def\d{\delta}

\begin{document}


\title{Relativistic effects for spin splitting of neutral particles: Upper bound and motional narrowing}

\author{Tihomir G. Tenev}
\affiliation{Department of Physics, Sofia University, 5 James Bourchier Blvd, Sofia 1164, Bulgaria}
\author{Nikolay V. Vitanov}
\affiliation{Department of Physics, Sofia University, 5 James Bourchier Blvd, Sofia 1164, Bulgaria}

\date{\today}

\begin{abstract}
We explore the properties of spin splitting for neutral particles possessing electric and magnetic dipole moments propagating in an electromagnetic field.
Two notable features of the spin splitting and the associated Larmor precession are found, which are consequences of special relativity.
First, we report the existence of an upper limit of spin splitting equal to twice the rest energy of the particle, and a corresponding upper limit for the Larmor precession frequency.
Second, we predict the noninvariance of the spin splitting and the corresponding Larmor frequency with respect to Lorentz boosts, which bears resemblance to the classical Doppler effect.
\end{abstract}

\pacs{03.67.Ac, 03.65.Pm, 37.10.Ty}
\keywords{Larmor precession, Dirac equation, special relativity, spin splitting, ion traps, Rydberg atoms}

\maketitle


The beginning of 20-th century physics has been marked by two major discoveries -- quantum theory and theory of relativity.
There was also a phenomenon that evaded every attempt for theoretical explanation for nearly twenty years -- the anomalous Zeeman effect.
It was finally explained by the hypothesis of spin and its quantum mechanical description by the Pauli matrices.
It is remarkable that these three topics were naturally united together by Dirac into his relativistic theory of the electron, which accounted naturally for the electron spin.
In fact, in his derivation the existence of spin appears as a consequence of the relativistic transformation of the quantum equation.
An early experimental success of the theory was the correct derivation of the size of the intrinsic magnetic moment of the electron.

Modified forms of the Dirac equation play the role of phenomenological equations describing more complex relativistic quantum dynamics.
Here we work with the extended Dirac equation~\cite{Thaller} for neutral particles possessing magnetic (MDM) and electric (EDM) dipole moments.
While magnetic dipole moments of elementary particles are experimentally well established facts, the hypothesis for electric dipole moments of elementary particles is still not experimentally proven despite being first proposed more than 50 years ago~\cite{Smith}.
We include it in our model for two reasons.
First, it is a prediction of the Standard Model and all its generalizations.
Second, we want to stress that the effects we consider do not depend on whether just a magnetic or electric field is present or a combination of both; they are valid in all these cases.

The basic experimental manifestation of MDM and EDM and the associated spin is the phenomenon of Larmor precession which stems from the underlying spin splitting of the energy levels of the stationary problem and which is measured in magnetic resonance.
We emphasize that most experiments and essentially all commercial devices relying on it perform measurements of particles moving far below relativistic speeds.

The relativistic theory predicts some unusual effects derived from just two postulates:
  (i) the constancy of speed of light for all inertial observers, and
 (ii) the invariance of the laws of nature with respect to different inertial observers.
Some of the well-known effects stemming from them are: (i) time dilatation; (ii) the Lorentz transformations; (iii) the existence of upper speed of propagation --- the speed of light, and (iv) the relationship between energy, momentum and rest mass $E^2=(pc)^2+(mc^2)^2$.
Another one is the relativistic Doppler effect which deviated from the nonrelativistic one because of time dilatation.

Here we predict the existence of two effects for spin splitting and Larmor precession, which we show to be consequences of special relativity.
The first effect is the existence of upper limit for spin splitting and associated quantities.
The second one is the noninvariance of spin splitting and associated quantities with respect to Lorentz boosts.

We consider a 1D model of a neutral relativistic particle with MDM $\mu$ and EDM $d$, which is placed in external static magnetic and/or electric fields directed along the propagation direction $x$ of the particle.
The Hamiltonian in SI units has the form
\begin{align}
\hat{H} &= c\hat{\alpha}_x \hat{p}_x + \hat{\beta}mc^2 + d\left(\i\hat{\beta}\hat{\alpha}_x B_x c + 2\hat{\beta}\hat{S}_x E_x  \right) \nonumber  \\ &+ \mu\left( \i \hat{\beta}\hat{\alpha}_x E_x/c - 2\hat{\beta}\hat{S}_x B_x\right) , \label{EQ:1DRelHam}
\end{align}
where $c$ is the speed of light, $m$ is the rest mass of the particle, $\hat{\alpha}_x$ and $\hat{\beta}$ are the Dirac matrices, and $\hat{S}_x$ is the x-component of the spin vector operator in relativistic theory $\vec{\hat{S}}=-\frac{1}{4}\i \vec{\alpha}\times\vec{\alpha}$, $\vec{E}$ is the electric field, $\vec{B}$ is the magnetic field.
For the sake of simplicity of notation, hereafter we shall drop the subscript $x$.
While the model \eqref{EQ:1DRelHam} can describe various neutral particles it is particularly suitable for the neutron as the one with the most reliable values of its parameters.

The four distinct eigenvalues of \eqref{EQ:1DRelHam} are
\begin{subequations}\label{EQ:EigenE}
\begin{align}
E_{\pm}^{\uparrow}   &= \pm \sqrt{\eta^2 + (mc^2 + \d )^2} ,  \\
E_{\pm}^{\downarrow} &= \pm \sqrt{\eta^2 + (mc^2 - \d )^2} ,
\end{align}
\end{subequations}
where we have introduced the notation
\begin{subequations}\label{EQ:defs}
\begin{align}
\eta&=\sqrt{(cp )^2 + (c\pi )^2},\\
\pi &= d B  + \mu E /c^2,\\
\d  &= d E - \mu B.
\end{align}
\end{subequations}
Unless the electric and magnetic fields are extremely strong (large $E$ and $B$) or the particles are very cold (small $p$),
 the term $(c\pi)^2$ is very small compared to the kinetic term $(cp)^2$,
 and the quantity $\eta$ is essentially proportional to the particle momentum $p$.

The size of the spin splitting is identical for positive and negative energy states, $\splitting = E_{+}^{\uparrow}-E_{+}^{\downarrow} = -(E_{-}^{\uparrow}-E_{-}^{\downarrow})$, given as
\begin{equation}
\splitting = \sqrt{\eta^2 + (mc^2+\d )^2} - \sqrt{\eta^2 + (mc^2 - \d )^2}.  \label{EQ:RelSpinSplit}
\end{equation}
This relativistic expression is substantially different from the nonrelativistic one,
\begin{equation}
\splitting_{\text{nonrel}} = 2\d = 2(d E - \mu B).
\end{equation}

Because
\begin{equation}
\partial_\eta\splitting = - \frac{\eta}{E_{+}^{\uparrow}E_{+}^{\downarrow}}\splitting \leqq 0 ,\label{EQ:Derivative}
\end{equation}
we conclude that $\splitting$ decreases when $\eta$ increases.
Therefore, the maximum value of $\splitting(\eta)$ occurs for $\eta=0$,
\begin{equation}
\splitting_0 = \splitting(\eta=0) = |mc^2+\d | - |mc^2-\d | ,\label{EQ:E0}
\end{equation}
which has two different values depending on the interaction energy $\d$:
\begin{subequations}\label{EQ:DE}
\begin{eqnarray}
\d  < mc^2\quad &\Longrightarrow&\quad \splitting_0 = 2\d , \label{EQ:DE1} \\
\d  \geqq mc^2\quad &\Longrightarrow&\quad \splitting_0 = 2mc^2. \label{EQ:DE2}
\end{eqnarray}
\end{subequations}
Therefore for $\d  < mc^2$ the spin splitting $\splitting_0$ reduces to the well-known nonrelativistic expression \eqref{EQ:DE1}, which increases linearly with the static magnetic and/or electric fields.
However if the interaction energy $\d $ exceeds the rest energy of the particle $mc^2$, the maximum spin splitting $\splitting_0$
 acquires the constant value \eqref{EQ:DE2}, which also happens to be the size of the mass gap -- the energy difference between positive and negative energy solutions of the free Dirac equation.
We conclude that the model embodied in the Hamiltonian \eqref{EQ:1DRelHam} predicts the existence of \emph{an upper limit for the spin splitting} for any neutral relativistic particle given by
\begin{equation}
\splitting_{\text{max}}=2mc^2. \label{EQ:UpperLimit}
\end{equation}
This naturally requires the existence of an upper limit of the Larmor frequency of spin precession
\begin{equation}\label{EQ:max splitting}
\omega_{\text{max}}=\frac{\splitting_{\text{max}}}{\hbar} = \frac{2mc^2}{\hbar}\; .
\end{equation}
It is interesting to note that the wavelength \mbox{$\lambda_{\text{min}}=c/\omega_{\text{max}}$}
 is equal to half the reduced Compton wavelength for the particle \mbox{$\lambda_C=\hbar/mc$}.

\begin{figure}[tb]
\includegraphics[width=0.9\columnwidth]{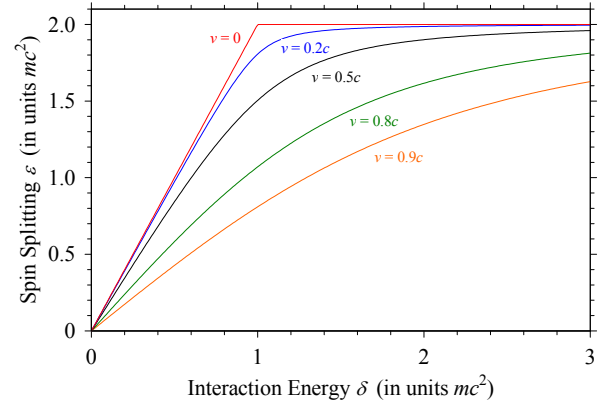}
\caption{(Color Online) Spin splitting vs. the interaction energy $\d$ for different velocities $v$ of a neutral relativistic particle.
The figure demonstrates the existence of an upper limit $\splitting_{\text{max}}=2mc^2$ of the spin splitting.
}\label{FIG:DEMax}
\end{figure}

%
We note that the existence of the upper limit of the spin splitting for the model \eqref{EQ:1DRelHam} is a consequence of the existence of the upper limit of the speed of light in special relativity. This is supported by the consideration of the nonrelativistic limit of Eq.~(\ref{EQ:UpperLimit}) by letting $c\rightarrow\infty$ we get $\splitting_{\text{max}}=2mc^2\rightarrow\infty$, which reproduces the limiting case of the nonrelativistic result.
Furthermore the existence of upper limit \eqref{EQ:max splitting} explains the disappearance of spin splitting for $m=0$ irrespective of the field strengths $E $ and $B $ and the EDM size $d_a$ and the MDM size $\mu_a$.

The existence of an upper limit for the spin splitting for the model \eqref{EQ:1DRelHam} is demonstrated in Fig.~\ref{FIG:DEMax} where the spin splitting is plotted as a function of the interaction energy $\d$ for several different velocities.
We use $\eta\approx cp$, and the relativistic relation $p=m\gamma v$ for a free particle, where $\gamma=1/\sqrt{1-(v/c)^2}$ is the Lorentz factor.
The dependence $\splitting(\d )$ for $\eta=0$ shows the expected linear increase of the spin splitting $\splitting$ with $\d $ up to the threshold value $\d =mc^2$, as described by Eq.~\eqref{EQ:DE}.
Above this threshold value for $\d$, the spin splitting achieves its upper value $\splitting_{\text{max}}$ and remains constant.
For finite velocities $\splitting(\d)$ increases monotonically but nonlinearly with $\d$.

\begin{figure}[tb]
\includegraphics[width=0.9\columnwidth]{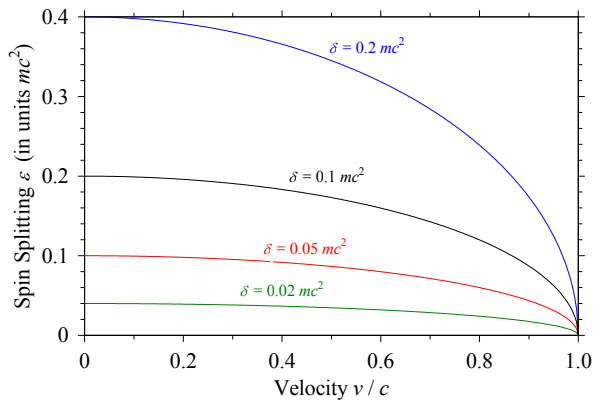}
\caption{(Color Online) Relativistic motional narrowing effect --- reduction of spin splitting when particle velocity increases within a relativistic model.
}\label{FIG:Motional}
\end{figure}

The qualitative behavior of the spin splitting $\splitting$ vs. $\eta$ can be deduced from Eq.~\eqref{EQ:Derivative} and is depicted on Fig.~\ref{FIG:Motional} for several different strengths of the interaction energy $\d$.
Because $\splitting(\eta)$ is a monotonically decreasing function of $\eta$, and because $\eta\approx cp$,
 this effect amounts to \emph{relativistic motional decrease of spin splitting}.
This effect is in stark contrast with the nonrelativistic result where the spin splitting is independent of the speed of the particle.

Simple estimates of the spin splitting relativistic motional narrowing can be obtained by considering the low speed and high speed limits of Eq.~\eqref{EQ:RelSpinSplit}.

\emph{\textbf{Low-speed limit.}}
When $\eta^2\ll (mc^2-\d )^2 < (mc^2+\d )^2$ we find from Eq.~\eqref{EQ:RelSpinSplit}:
\begin{align}
\splitting &= \splitting_0 - \frac{\splitting_0}{2|m^2c^4-\d ^2|} \eta^2 + O(\eta^4) .\label{EQ:LowDopler}
\end{align}
This quadratic departure of $\splitting$ from its maximum value $2mc^2$ is indeed observed in Fig.~\ref{FIG:Motional}.

The relativistic motional decrease of the spin splitting leads to decrease of the value of Larmor frequency of precession of spin around an external static field, which resembles the Doppler effect for emitted light.
We refer to this phenomenon as \emph{relativistic motional red shift} of the Larmor precession.
We use $\eta\approx cp$, and the relativistic relation $p=m\gamma v$ for a free particle, where $\gamma=1/\sqrt{1-(v/c)^2}$ is the Lorentz factor.
We find from Eq.~\eqref{EQ:LowDopler} that
\begin{equation}
\frac{\omega(v)}{\omega_0} = 1 - \frac{c^2(m\gamma v)^2}{2\left|(m^2c^4-\d^2\right|}.\label{EQ:FreqLowDoppler}
\end{equation}
The frequency change itself is given by
\begin{equation}
 \omega(v)-\omega_0=-\frac{c^2(m\gamma v)^2}{2\left|(m^2c^4-\d^2\right|}\omega_0.
\end{equation}
For comparison, the red-shift effect of the nonrelativistic Dopper effect is described by the expression $\omega/\omega_0=1-v/c$, and the relativistic one by $\omega/\omega_0=\gamma(1-v/c)$, where $\omega$ and $\omega_0$ are the angular frequencies of emitted light in the observer and emitter frames of reference, respectively.
The change of frequency of the nonrelativistic and relativistic Doppler effects depend only on the ratio $v/c$ and the difference between the two is due to the time-dilatation effect. In contrast the change of frequency in the \emph{relativistic motional red shift} depends on three factors: (i) the rest mass $m$ of the particle, (ii) the magnitude of the interaction energy $\d $, and (iii) the velocity $v$ through the term \mbox{$(\gamma v)^2$}.

\emph{\textbf{High-speed limit.}}
In the limit of near light speed ($v\to c$), we have $\eta^2\approx (cp)^2=(c\gamma mv)^2\to\infty$.
We find from Eq.~\eqref{EQ:RelSpinSplit} that
\begin{equation}
\splitting(\eta\to\infty) = 2\d \frac{mc^2}{\eta} + O(\eta^{-3}).
\end{equation}
It shows that the spin splitting $\splitting$ tends to zero inversely proportionally to $p$ because $\eta\approx cp$.
Three other features are notable: (i)  $\splitting(v\to c)$ does not depend on $\splitting_0$; (ii) $\splitting(v\to c)$ depends quadratically on the interaction energy $\d $, and (iii) $\splitting(v\sim c)$ does not depend on the rest mass $m$ because $\eta\approx mc\gamma v$.

The physical explanation of the relativistic motional narrowing of spin splitting is easy to trace to the specific type of noninvariance of momentum and energy in special relativity.
For example, for a free particle they are related by $E^2-(pc)^2=(mc^2)^2$, the only invariant quantity being $mc^2$.
This explains both the presence of negative energies and the square-root dependence in the eigenenergies \eqref{EQ:EigenE}.
The square-root dependence combined with the splitting between $E^{\uparrow}$ and $E^{\downarrow}$ gives the essential ingredients of the relativistic motional narrowing and red-shift effects.
In contrast, for Galilean invariant systems the momentum parts of spin-up and spin-down eigenenergies do not contribute to the spin splitting, thereby leaving it invariant with respect to Galilean transformations.

The two predicted effects --- upper bound of spin splitting and motional narrowing (or Larmor red shift) --- can be observed experimentally using the existing technology.
It is unlikely to be able to probe the absolute limit for spin splitting of neutral particles $\splitting_{\text{max}}=2mc^2$ by conventional experiments due to the large energy associated with the rest-mass and the relatively small spin splittings that can be achieved.
However, it may be possible to probe this limit in emulation of spin splitting of neutral particles with trapped ions, for example by using our recent proposal \cite{Tenev} for emulation of EDM of neutral relativistic particles moving in electrostatic fields.
This proposal is easily adaptable to emulation of MDM in static field by identifying the Rabi frequency of the carrier interaction with MDM-magnetic field coupling as $2\hbar\Omega_j^{(1)}=-\mu_a B_j$.
Trapped ions have proved to be a suitable physical platform for simulation of relativistic effects in recent years \cite{Lamata2007,Bermudez2007a,Bermudez2007b,Casanova2010,Casanova2011};
 some effects have been already demonstrated experimentally \cite{Gerritsma2010,Gerritsma2011}.

The relativistic motional narrowing effect may be emulated with trapped ions using the same experimental setup as for the existence of upper limit of spin splitting.
However, it may also be possible to detect it with conventional magnetic resonance techniques.
For that purpose one will need to measure Larmor precession frequency along the same magnetic field for neutrons propagating with different speeds sufficiently greater than $0.1c$.
Similar technique may be used to test the relativistic motional narrowing effect using Rydberg atoms \cite{RydbergAtoms}.
They are suitable for this experiment thanks to their large induced electric dipole moment combined with neutral charge.

In conclusion, by examining a relativistic model for neutral particles we have predicted the existence of upper limit for the spin splitting equal to twice the rest energy, $\splitting_{\text{max}}=2mc^2$.
This upper limit is unique for every particle and depends on its rest mass.
This result is in contrast to nonrelativistic models where such limit does not exist.
We have argued that its existence is a consequence of the existence of an upper limit of speed of propagation in relativistic theory.
We have shown that within the considered model the spin splitting $\splitting$ depends on particle momentum and that for higher velocities the spin splitting decreases, as illustrated in Fig.~(\ref{FIG:Motional}).
We have referred to this effect as \emph{relativistic motional narrowing effect} and have explained its origin in the noninvariance of the energy-momentum relation with respect to Lorentz boosts in relativistic theory.
The predicted effects can be observed experimentally in at least three different physical systems: neutrons, emulation in trapped ions, and Rydberg atoms.

We thank Peter A. Ivanov for fruitful discussions.
This work has been supported by the EU COST project IOTA and the Bulgarian NSF grants D002-90/08 and DMU03/107.

%

\end{document}